\documentclass{article}
\usepackage{amsfonts,amsmath,amsthm,amstext,epsfig,mathbbol}
\usepackage{subfigure}
\RequirePackage{ifthen}
\RequirePackage[usenames]{pstcol}
\RequirePackage{pst-node}
\RequirePackage{pst-plot}
\RequirePackage{pst-coil}
\RequirePackage{multido}
\RequirePackage{pst-3d}
\RequirePackage{color}
\RequirePackage{calc}
\newcommand{\SQRTwo}{0.717}

\newcommand{\StackTwoLabels}[2]{%
   \renewcommand{\arraystretch}{0.75}%
   \begin{array}{c}#1\\ #2 \end{array}%
   \renewcommand{\arraystretch}{1.333}}

\newcommand{\MediumScale}{0.6}

\newlength{\MediumStateDiameter}
\newlength{\SmallStateDiameter}
\newlength{\LargeStateDiameter}
\newlength{\VerySmallStateDiameter}
\newlength{\StateLineWidth}        
\newcommand{\StateLineStyle}{solid} 
\newcommand{\StateLineColor}{black}
\newif\ifStateLineDbl \StateLineDblfalse 
\newcommand{\StateLineDblCoef}{0.6} 
\newcommand{\StateLineDblSep}{0.4} 
\newcommand{\VSStateLineCoef}{.6} 
\newcommand{\StateFillStatus}{solid} 
\newcommand{\StateFillColor}{white}
\newcommand{\StateLabelColor}{black}
\newcommand{\StateLabelScale}{1.7}
\newcommand{\DimStateLineStyle}{solid} 
\newcommand{\DimStateLineCoef}{1} %
\newcommand{\DimStateLineColor}{gray}
\newcommand{\DimStateLabelColor}{gray}
\newcommand{\DimStateFillColor}{white}
\newlength{\EdgeLineWidth}
\newcommand{\EdgeLineStyle}{solid}
\newif\ifEdgeLineDbl \EdgeLineDblfalse 
\newcommand{\EdgeLineDblCoef}{0.5} 
\newcommand{\EdgeLineDblSep}{0.6} 
\newcommand{\EdgeLineColor}{black}
\newlength{\EdgeArrowWidth}
\newlength{\EdgeDblArrowWidth}
\newcommand{\EdgeArrowLengthCoef}{1.4}
\newcommand{\EdgeDblArrowLengthCoef}{1.7}
\newcommand{\EdgeArrowInset}{0.1}
\newcommand{\EdgeArrowStyle}{->}
\newcommand{\EdgeRevArrowStyle}{<-}
\newcommand{\EdgeLineBorderCoef}{2}
\newcommand{\EdgeLineBorderColor}{white}
\newcommand{\EdgeLabelColor}{black}
\newcommand{\EdgeLabelScale}{1.7}
\newcommand{\DimEdgeLineCoef}{1.2} 
\newcommand{\DimEdgeLineStyle}{solid} 
\newcommand{\DimEdgeLineColor}{gray}
\newcommand{\DimEdgeLabelColor}{gray}
\newlength{\ZZSize}
\newcommand{\ZZShape}{0.5}
\newcommand{\ZZLineWidth}{1.7}
\newcommand{\TransLabelZZCoef}{0.6}
\newlength{\EdgeOffset}
\newcommand{\ForthBackEdgeOffset}{5}
\newcommand{\VaucArcAngle}{15}
\newcommand{\VaucArcCurvature}{0.8}
\newlength{\VaucArcOffset}
\newcommand{\VaucLArcAngle}{30}
\newcommand{\VaucLArcCurvature}{0.8}
\newlength{\LoopOffset}
\newlength{\LoopVarOffset}
\newcommand{\LoopAngle}{30}
\newcommand{\CLoopAngle}{22}
\newcommand{\LoopVarAngle}{28}
\newcommand{\LoopOnMediumState}{7}
\newcommand{\LoopOnSmallState}{9.6} 
\newcommand{\LoopOnLargeState}{5.8}

\newcommand{\CLoopOnMediumState}{8}
\newcommand{\CLoopOnSmallState}{12}
\newcommand{\CLoopOnLargeState}{6}

\newlength{\TransLabelSep}
\newcommand{\EdgeLabelPosit}{.45}\newcommand{\EdgeLabelRevPosit}{.55}
\newcommand{\ArcLabelPosit}{.4}
\newcommand{\LArcLabelPosit}{.4}
\newcommand{\LoopLabelPosit}{.25}
\newcommand{\CLoopLabelPosit}{.25}
\newcommand{\InitStateLabelPosit}{.1}
\newcommand{\FinalStateLabelPosit}{.9}
\newcommand{\ArrowOnStateCoef}{}
\newcommand{\ArrowOnMediumState}{1.5}
\newcommand{\ArrowOnSmallState}{1.7} 
\newcommand{\ArrowOnLargeState}{1.3}
\newcommand{\ArrowOnVerySmallState}{5} 
\newlength{\VertShiftH} 
\newlength{\VertShiftD} 
\newlength{\VertShift}
\newif\ifVCFrame
\newcommand{\HideFrame}{\VCFramefalse}

\newif\ifVCGrid
\newcommand{\HideGrid}{\VCGridfalse}

\newif\ifVCRigidLabel
\newcommand{\RigidLabel}{\VCRigidLabeltrue}

\newif\ifVCStateLabelBaseLine

\psset{unit=1cm}
\newpsstyle{VaucFrameStyle}{arrows=-,framesep=0pt,%
                     linewidth=0.6pt,linecolor=black,%
                                         linestyle=solid,%
                                         doubleline=false,%
                                         fillcolor=white,fillstyle=none,%
                                         cornersize=relative,framearc=0}
\newcommand{\FrameStyle}{\psset{style=VaucFrameStyle}}
\newpsstyle{VaucGridStyle}{%
      gridwidth=0.6pt,griddots=10,subgriddiv=1,%
          gridlabels=7pt}
\newcommand{\GridStyle}{\psset{style=VaucGridStyle}}
\newenvironment{VCPicture}[2][.5]%
  {\settoheight{\VertShiftH}{$\{$}%
   \settodepth{\VertShiftD}{$\{$}%
   \setlength{\VertShift}{.5\VertShiftD-.5\VertShiftH}%
   \begin{pspicture}[#1]#2%
   \ifVCFrame \FrameStyle \psframe#2\fi%
   \ifVCGrid \FrameStyle\GridStyle \psgrid#2\fi}%
  {\RstState\RstEdge%
   \end{pspicture}}
\newcommand{\VCScale}{}
\newcommand{\FixVCScale}[1]{\renewcommand{\VCScale}{#1}}

\newcommand{\MediumPicture}{\FixVCScale{\MediumScale}}

\newcommand{\VCDirectory}{}
\newcommand{\SetVCDirectory}[1]{\renewcommand{\VCDirectory}{#1}}
\newif\ifVCName
\newcommand{\HideName}{\VCNamefalse}

\newcommand{\VCDraw}[2][\VCGridScale]{%
\psset{unit=#1cm}%
\ifVCName\makebox[0pt][r]{\fbox{{\scriptsize #2}}}\fi%
\scalebox{\VCScale}{#2}%
\psset{unit=1cm}}

\newcommand{\VCPut}[3][0]{\rput{#1}#2{#3}}%
\newlength{\StateLineWid}
\newcommand{\StateLineSty}{\StateLineStyle} 
\newcommand{\StateLineCol}{\StateLineColor}

\newcommand{\StateFillCol}{\StateFillColor}
\newcommand{\StateFillSta}{\StateFillStatus} 
\newcommand{\StateLabelSca}{1}
\newcommand{\StateLabelCol}{\StateLabelColor}
\newcommand{\StateDimen}{outer}
\newcommand{\StateDblDimen}{middle}
\newcommand{\VCIFflag}{}
\newcommand{\PlainState}%
  {\renewcommand{\VCIFflag}{0}}
\newcommand{\FullState}%
  {\renewcommand{\VCIFflag}{2}}

\newif\ifVCShowState

\newcommand{\ShowState}{\VCShowStatetrue}
\ShowState 
\newpsstyle{VaucStateStyle}{framesep=0pt,%
         linewidth=\StateLineWid,linecolor=\StateLineCol,%
         linestyle=\StateLineSty,doubleline=false,%
         fillcolor=\StateFillCol,fillstyle=\StateFillSta,%
         border=0pt,dimen=\StateDimen,%
         cornersize=relative,framearc=1,framesep=0pt}
\newpsstyle{VaucStateDblStyle}{framesep=0pt,%
         linewidth=\StateLineDblCoef\StateLineWid,linecolor=\StateLineCol,%
         linestyle=\StateLineSty,doubleline=true,doublesep=\StateLineDblSep\StateLineWid,%
         fillcolor=\StateFillCol,fillstyle=\StateFillSta,%
         border=0pt,dimen=\StateDblDimen,%
         cornersize=relative,framearc=1,framesep=0pt}
\newpsstyle{VaucHiddenStateStyle}{framesep=0pt,%
         linewidth=\StateLineWid,linecolor=\StateLineCol,%
         linestyle=none,%
         fillcolor=\StateFillCol,fillstyle=none,%
         border=0pt,dimen=outer,%
         cornersize=relative,framearc=1,framesep=0pt}
\newcommand{\StateStyle}{%
   \ifVCShowState%
         \ifStateLineDbl\psset{style=VaucStateDblStyle}\else\psset{style=VaucStateStyle}\fi%
   \else\psset{style=VaucHiddenStateStyle}\fi}
\newcommand{\VaucStateRBLabel}[1]{%
    \textcolor{\StateLabelCol}{\scalebox{\StateLabelSca}{\scalebox{\StateLabelScale}{\rput[B]{0}(0,\VertShift){$ #1 $}}}}}%
\newcommand{\VaucStateLabel}[1]%
    {\ifVCShowState%
        \ifVCRigidLabel%
           \ifVCStateLabelBaseLine%
                 \textcolor{\StateLabelCol}{\scalebox{\StateLabelSca}{\scalebox{\StateLabelScale}{\rput[B]{*0}(0,\VertShift){$ #1 $}}}}%
           \else
                 \textcolor{\StateLabelCol}{\scalebox{\StateLabelSca}{\scalebox{\StateLabelScale}{\rput{*0}(0,0){$ #1 $}}}}%
           \fi
        \else
                 \textcolor{\StateLabelCol}{\scalebox{\StateLabelSca}{\scalebox{\StateLabelScale}{$ #1 $}}}%
        \fi
     \else%
                 \textcolor{white}{\scalebox{\StateLabelSca}{\scalebox{\StateLabelScale}{$ #1 $}}}%
     \fi}
\newcommand{\VCPutStateLabel}[2]%
    {\rput#1{\scalebox{\StateLabelSca}{$ #2 $}}}%
\newcommand{\ChgStateLineStyle}[1]{\renewcommand{\StateLineSty}{#1}}
\newcommand{\RstStateLineStyle}{\ChgStateLineStyle{\StateLineStyle}}
\newcommand{\SetStateLineStyle}[1]%
   {\renewcommand{\StateLineStyle}{#1}\RstStateLineStyle}%

\newcommand{\ChgStateLineWidth}[1]{\setlength{\StateLineWid}{#1\StateLineWidth}}%
\newcommand{\RstStateLineWidth}{\ChgStateLineWidth{1}}%
\newcommand{\SetStateLineWidth}[1]
   {\setlength{\StateLineWidth}{#1}\RstStateLineWidth}
\newcommand{\ChgStateLineColor}[1]{\renewcommand{\StateLineCol}{#1}}
\newcommand{\RstStateLineColor}{\ChgStateLineColor{\StateLineColor}}
\newcommand{\SetStateLineColor}[1]%
   {\renewcommand{\StateLineColor}{#1}\RstStateLineColor}
\newcommand{\ChgStateFillStatus}[1]{\renewcommand{\StateFillSta}{#1}}
\newcommand{\RstStateFillStatus}{\ChgStateFillStatus{\StateFillStatus}}
\newcommand{\SetStateFillStatus}[1]%
    {\renewcommand{\StateFillStatus}{#1}\RstStateFillStatus}
\newcommand{\ChgStateFillColor}[1]{\renewcommand{\StateFillCol}{#1}}
\newcommand{\RstStateFillColor}{\ChgStateFillColor{\StateFillColor}}
\newcommand{\SetStateFillColor}[1]%
    {\renewcommand{\StateFillColor}{#1}\RstStateFillColor}%
\newcommand{\ChgStateLabelColor}[1]{\renewcommand{\StateLabelCol}{#1}}
\newcommand{\RstStateLabelColor}{\ChgStateLabelColor{\StateLabelColor}}
\newcommand{\SetStateLabelColor}[1]%
    {\renewcommand{\StateLabelCol}{#1}\RstStateLabelColor}
\newcommand{\ChgStateLabelScale}[1]{\renewcommand{\StateLabelSca}{#1}}
\newcommand{\RstStateLabelScale}{\ChgStateLabelScale{1}}
\newcommand{\SetStateLabelScale}[1]%
   {\renewcommand{\StateLabelScale}{#1}\RstStateLabelScale}
\newcommand{\FixStateLineDouble}[2]{%
    \renewcommand{\StateLineDblCoef}{#1}%
    \renewcommand{\StateLineDblSep}{#2}}
\newcommand{\FixDimState}[5]{%
    \renewcommand{\DimStateLineStyle}{#1}%
    \renewcommand{\DimStateLineCoef}{#3}%
    \renewcommand{\DimStateLineColor}{#2}%
    \renewcommand{\DimStateLabelColor}{#4}%
    \renewcommand{\DimStateFillColor}{#5}}%
\newcommand{\RstState}{%
   \RstStateLineStyle\RstStateLineWidth%
   \RstStateLineColor%
   \RstStateFillStatus\RstStateFillColor%
   \RstStateLabelColor\RstStateLabelScale}%
\newlength{\StateDiam}
\newlength{\VaucAOS}\newlength{\VaucAOSdiag}
\newcommand{\StateSizeFlag}{}
\newcommand{\SetAOS}{%
   \setlength{\VaucAOS}{\ArrowOnStateCoef\StateDiam}%
   \setlength{\VaucAOSdiag}{\SQRTwo\VaucAOS}}
\newlength{\VariableStateIntDiam}
\newlength{\VariableStateWidth}
\newlength{\VariableStateITPos}
\newcommand{\SetStateIntDiam}{%
   \setlength{\VariableStateIntDiam}{\StateDiam}%
   \addtolength{\VariableStateIntDiam}{-2\StateLineWid}%
}%
\newcommand{\LoopSize}{}\newcommand{\LoopSi}{}
\newcommand{\LoopVarSize}{}\newcommand{\LoopVarSi}{}
\newcommand{\CLoopSize}{}\newcommand{\CLoopSi}{}
\newcommand{\ChgLoopSize}[1]{\renewcommand{\LoopSi}{#1}}
\newcommand{\RstLoopSize}{\ChgLoopSize{\LoopSize}}
\newcommand{\SetLoopSize}[1]%
   {\renewcommand{\LoopSize}{#1}\RstLoopSize}
\newcommand{\ChgCLoopSize}[1]{\renewcommand{\CLoopSi}{#1}}
\newcommand{\RstCLoopSize}{\ChgCLoopSize{\CLoopSize}}
\newcommand{\SetCLoopSize}[1]%
   {\renewcommand{\CLoopSize}{#1}\RstCLoopSize}
\newcommand{\ChgLoopVarSize}[1]{\renewcommand{\LoopVarSi}{#1}}
\newcommand{\RstLoopVarSize}{\ChgLoopVarSize{\LoopVarSize}}
\newcommand{\SetLoopVarSize}[1]%
   {\renewcommand{\LoopVarSize}{#1}\RstLoopVarSize}
\newcommand{\SetStateDiam}[4]{%
   \setlength{\StateDiam}{#1}%
   \renewcommand{\ArrowOnStateCoef}{#2}%
   \SetLoopSize{#3}%
   \SetLoopVarSize{#3}%
   \SetCLoopSize{#4}%
   \SetAOS\SetStateIntDiam}
\newcommand{\FixStateDiameter}[1]
   {\setlength{\StateDiam}{#1}\SetStateIntDiam \SetAOS}
\newcommand{\MediumState}%
   {\SetStateDiam{\MediumStateDiameter}{\ArrowOnMediumState}%
         {\LoopOnMediumState}{\CLoopOnMediumState}%
                  \renewcommand{\StateSizeFlag}{0}}
\newcommand{\SmallState}%
   {\SetStateDiam{\SmallStateDiameter}{\ArrowOnSmallState}%
         {\LoopOnSmallState}{\CLoopOnSmallState}%
                  \renewcommand{\StateSizeFlag}{1}}
\newcommand{\LargeState}%
   {\SetStateDiam{\LargeStateDiameter}{\ArrowOnLargeState}%
         {\LoopOnLargeState}{\CLoopOnLargeState}%
                  \renewcommand{\StateSizeFlag}{2}}
\newcommand{\RstStateSize}%
  {\ifthenelse{\equal{\StateSizeFlag}{0}}%
              {\MediumState}%
              {\ifthenelse{\equal{\StateSizeFlag}{1}}%
                              {\SmallState}{\LargeState}}}%
\newcommand{\VaucState}[3][{}]%
   {\rput#2{%
      \Cnode[radius=.5\StateDiam](0,0){#3}%
          \ifVCShowState%
      \nput[labelsep=-.5\StateDiam]{0}{#3}%
        {\makebox[0pt]{\VaucStateLabel{#1}}}%
      \fi
      \ifthenelse{\equal{\VCIFflag}{0}}{}{%
        \pnode(-\VaucAOS,0){#3w}\pnode(\VaucAOS,0){#3e}%
        \pnode(0,\VaucAOS){#3n}\pnode(0,-\VaucAOS){#3s}%
           \ifthenelse{\equal{\VCIFflag}{1}}{}{%
          \pnode(-\VaucAOSdiag,\VaucAOSdiag){#3nw}%
           \pnode(\VaucAOSdiag,\VaucAOSdiag){#3ne}%
           \pnode(-\VaucAOSdiag,-\VaucAOSdiag){#3sw}%
           \pnode(\VaucAOSdiag,-\VaucAOSdiag){#3se}%
                   }%
            }%
     }%
}
\newcommand{\State}[3][{}]{\StateStyle\VaucState[#1]{#2}{#3}}
\newcommand{\FinalState}[3][{}]%
   {\psset{style=VaucStateDblStyle}\VaucState[#1]{#2}{#3}}
\newcommand{\VSState}[2]%
    {\renewcommand{\ArrowOnStateCoef}{\ArrowOnVerySmallState}%
         \FixStateDiameter{\VerySmallStateDiameter}%
     \ChgStateLineWidth{\VSStateLineCoef}%
         \State{#1}{#2}%
         \RstStateLineWidth\RstStateSize}
\newlength{\ExtraSpace}
\newcommand{\StateVar}[3][]%
 {\StateStyle %
  \settowidth{\VariableStateWidth}{\scalebox{\StateLabelSca}{\scalebox{\StateLabelScale}{$#1$}}}%
  \addtolength{\VariableStateWidth}{\ExtraSpace}
  \ifthenelse{\lengthtest{\VariableStateWidth < \VariableStateIntDiam}}%
        {\setlength{\VariableStateWidth}{\VariableStateIntDiam}}{}%
  \setlength{\VariableStateITPos}{\ArrowOnStateCoef\StateDiam}%
  \addtolength{\VariableStateITPos}{0.5\VariableStateWidth}%
  \addtolength{\VariableStateITPos}{-0.5\StateDiam}%
  \rput#2{\pnode(\VariableStateITPos,0){#3e}%
          \pnode(-\VariableStateITPos,0){#3w}%
          \pnode(0,\ArrowOnStateCoef\StateDiam){#3n}%
          \pnode(0,-\ArrowOnStateCoef\StateDiam){#3s}}%
  \rput#2{\rnode{#3}{\psframebox{\protect\rule[-.5\VariableStateIntDiam]{0pt}{\VariableStateIntDiam}\protect\rule{\VariableStateWidth}{0pt}}}}
  \rput#2{\VaucStateRBLabel{#1}}%
}%
\newcommand{\StateSqr}[3][]%
 {\StateStyle%
  \settowidth{\VariableStateWidth}{\scalebox{\StateLabelSca}{\scalebox{\StateLabelScale}{$#1$}}}%
  \addtolength{\VariableStateWidth}{\ExtraSpace}
  \ifthenelse{\lengthtest{\VariableStateWidth < \VariableStateIntDiam}}%
        {\setlength{\VariableStateWidth}{\VariableStateIntDiam}}{}%
  \setlength{\VariableStateITPos}{\ArrowOnStateCoef\StateDiam}%
  \addtolength{\VariableStateITPos}{0.5\VariableStateWidth}%
  \addtolength{\VariableStateITPos}{-0.5\StateDiam}%
  \rput#2{\pnode(\VariableStateITPos,0){#3e}%
          \pnode(-\VariableStateITPos,0){#3w}%
          \pnode(0,\ArrowOnStateCoef\StateDiam){#3n}%
          \pnode(0,-\ArrowOnStateCoef\StateDiam){#3s}}%
  \rput#2{\rnode{#3}{\psframebox[framearc=0]{\protect\rule[-.5\VariableStateIntDiam]{0pt}{\VariableStateIntDiam}\protect\rule{\VariableStateWidth}{0pt}}}}
  \rput#2{\VaucStateRBLabel{#1}}%
}%

\newlength{\EdgeLineWid}
\newcommand{\EdgeLineSty}{\EdgeLineStyle}
\newcommand{\EdgeLineCol}{\EdgeLineColor}
\newcommand{\EdgeLabelSca}{1}
\newcommand{\EdgeLabelCol}{\EdgeLabelColor}
\newlength{\EdgeArrowSZDim}
\newcommand{\EdgeArrowSZNum}{\EdgeArrowLengthCoef}
\newcommand{\EdgeArrowSty}{\EdgeArrowStyle}
\newcommand{\EdgeArrowIns}{\EdgeArrowInset}
\newlength{\EdgeLineBord}
\newlength{\ZZSiZ}
\newcommand{\ZZLineWid}{\ZZLineWidth}
\newlength{\EdgeOff}
\newcommand{\VaucArcAng}{\VaucArcAngle}
\newcommand{\VaucLArcAng}{\VaucLArcAngle}
\newlength{\VaucArcOff}
\newcommand{\VaucArcCurv}{\VaucArcCurvature}
\newcommand{\VaucLArcCurv}{\VaucLArcCurvature}
\newcommand{\LoopAng}{\LoopAngle}
\newcommand{\CLoopAng}{\CLoopAngle}
\newcommand{\LoopVarAng}{\LoopVarAngle}
\newlength{\LoopOff}
\newlength{\LoopVarOff}
\newlength{\TransLabelSP}
\newcommand{\EdgeLabelPos}{\EdgeLabelPosit}
\newcommand{\ArcLabelPos}{\ArcLabelPosit}
\newcommand{\LArcLabelPos}{\LArcLabelPosit}
\newcommand{\LoopLabelPos}{\LoopLabelPosit}
\newcommand{\CLoopLabelPos}{\CLoopLabelPosit}
\newcommand{\InitStateLabelPos}{\InitStateLabelPosit}
\newcommand{\FinalStateLabelPos}{\FinalStateLabelPosit}
\newpsstyle{VaucEdgeStyle}%
    {arrows=\EdgeArrowSty,arrowsize=\EdgeArrowSZDim,arrowlength=\EdgeArrowSZNum,%
         arrowinset=\EdgeArrowIns,%
     linewidth=\EdgeLineWid,linecolor=\EdgeLineCol,linestyle=\EdgeLineSty,%
     doubleline=false,%
         bordercolor=\EdgeLineBorderColor,border=\EdgeLineBord,%
     fillstyle=none,offset=\EdgeOff,%
     labelsep=\TransLabelSP,nodesep=0pt}
\newpsstyle{VaucEdgeDblStyle}%
    {arrows=\EdgeArrowSty,arrowsize=\EdgeArrowSZDim,arrowlength=\EdgeArrowSZNum,%
         arrowinset=\EdgeArrowIns,%
     linewidth=\EdgeLineDblCoef\EdgeLineWid,linecolor=\EdgeLineCol,linestyle=\EdgeLineSty,%
     doubleline=true,doublesep=\EdgeLineDblSep\EdgeLineWid,%
         bordercolor=\EdgeLineBorderColor,border=\EdgeLineBord,%
     fillstyle=none,offset=\EdgeOff,%
     labelsep=\TransLabelSP,nodesep=0pt}
\newpsstyle{VaucArcR}{ncurv=\VaucArcCurv,arcangle=-\VaucArcAng,%
     labelsep=\TransLabelSP,offset=-\VaucArcOff}
\newpsstyle{VaucArcL}{ncurv=\VaucArcCurv,arcangle=\VaucArcAng,%
     labelsep=\TransLabelSP,offset=\VaucArcOff}
\newpsstyle{VaucLArcR}{ncurv=\VaucLArcCurv,arcangle=-\VaucLArcAng,%
     labelsep=\TransLabelSP,offset=-\VaucArcOff}
\newpsstyle{VaucLArcL}{ncurv=\VaucLArcCurv,arcangle=\VaucLArcAng,%
     labelsep=\TransLabelSP,offset=\VaucArcOff}
\newpsstyle{VaucZigzagStyle}%
   {linewidth=\ZZLineWid\EdgeLineWid,%
    labelsep=\TransLabelSP,nodesep=0pt,%
    coilwidth=1.2\ZZSiZ,coilarmA=0.1\ZZSiZ,%
    coilarmB=0.3\ZZSiZ,coilheight=\ZZShape,linearc=1.6pt}
\newcommand{\EdgeStyle}{\ifEdgeLineDbl\psset{style=VaucEdgeDblStyle}%
        \else\psset{style=VaucEdgeStyle}\fi}
\newcommand{\ZigzagStyle}%
   {\addtolength{\TransLabelSP}{\TransLabelZZCoef\ZZSiZ}%
    \psset{style=VaucZigzagStyle}%
        \addtolength{\TransLabelSP}{-\TransLabelZZCoef\ZZSiZ}%
        }
\newcommand{\ChgEdgeOffset}[1]{\setlength{\EdgeOff}{#1}}
\newcommand{\RstEdgeOffset}{\ChgEdgeOffset{\EdgeOffset}}
\newcommand{\SetEdgeOffset}[1]%
   {\setlength{\EdgeOffset}{#1}\RstEdgeOffset}
\newcommand{\ForthBackOffset}{%
   \setlength{\EdgeOff}{\ForthBackEdgeOffset\EdgeLineWid}}
\newcommand{\ChgArcAngle}[1]{\renewcommand{\VaucArcAng}{#1}}
\newcommand{\RstArcAngle}{\ChgArcAngle{\VaucArcAngle}}
\newcommand{\SetArcAngle}[1]%
   {\renewcommand{\VaucArcAngle}{#1}\RstArcAngle}
\newcommand{\ChgLArcAngle}[1]{\renewcommand{\VaucLArcAng}{#1}}
\newcommand{\RstLArcAngle}{\ChgLArcAngle{\VaucLArcAngle}}
\newcommand{\SetLArcAngle}[1]%
   {\renewcommand{\VaucLArcAngle}{#1}\RstLArcAngle}
\newcommand{\ChgArcCurvature}[1]{\renewcommand{\VaucArcCurv}{#1}}
\newcommand{\RstArcCurvature}{\ChgArcCurvature{\VaucArcCurvature}}
\newcommand{\SetArcCurvature}[1]%
   {\renewcommand{\VaucArcCurvature}{#1}\RstArcCurvature}
\newcommand{\ChgLArcCurvature}[1]{\renewcommand{\VaucLArcCurv}{#1}}
\newcommand{\RstLArcCurvature}{\ChgLArcCurvature{\VaucLArcCurvature}}
\newcommand{\SetLArcCurvature}[1]%
   {\renewcommand{\VaucLArcCurvature}{#1}\RstLArcCurvature}

\newcommand{\RstArcOffset}{\setlength{\VaucArcOff}{\VaucArcOffset}}
\newcommand{\SetArcOffset}[1]%
   {\renewcommand{\VaucArcOffset}{#1}\RstArcOffset}

\newcommand{\RstLoopOffset}{\setlength{\LoopOff}{\LoopOffset}}
\newcommand{\SetLoopOffset}[1]%
   {\renewcommand{\LoopOffset}{#1}\RstLoopOffset}
\newcommand{\ChgLoopAngle}[1]{\renewcommand{\LoopAng}{#1}}
\newcommand{\RstLoopAngle}{\ChgLoopAngle{\LoopAngle}}
\newcommand{\SetLoopAngle}[1]%
   {\renewcommand{\LoopAngle}{#1}\RstLoopAngle}
\newcommand{\ChgCLoopAngle}[1]{\renewcommand{\CLoopAng}{#1}}
\newcommand{\RstCLoopAngle}{\ChgCLoopAngle{\CLoopAngle}}
\newcommand{\SetCLoopAngle}[1]%
   {\renewcommand{\CLoopAngle}{#1}\RstCLoopAngle}
\newcommand{\ChgEdgeLineColor}[1]{\renewcommand{\EdgeLineCol}{#1}}
\newcommand{\RstEdgeLineColor}{\ChgEdgeLineColor{\EdgeLineColor}}
\newcommand{\SetEdgeLineColor}[1]%
   {\renewcommand{\EdgeLineColor}{#1}\RstEdgeLineColor}
\newcommand{\ChgEdgeLineStyle}[1]{\renewcommand{\EdgeLineSty}{#1}}  
\newcommand{\RstEdgeLineStyle}{\ChgEdgeLineStyle{\EdgeLineStyle}}
\newcommand{\SetEdgeLineStyle}[1]%
   {\renewcommand{\EdgeLineStyle}{#1}\RstEdgeLineStyle}
\newcommand{\ChgEdgeLineWidth}[1]
   {\setlength{\EdgeLineWid}{#1\EdgeLineWidth}}
\newcommand{\RstEdgeLineWidth}{\ChgEdgeLineWidth{1}}
\newcommand{\SetEdgeLineWidth}[1]
   {\setlength{\EdgeLineWidth}{#1}\RstEdgeLineWidth}
\newcommand{\EdgeLineDouble}%
        {\EdgeLineDbltrue%
    \ChgEdgeArrowWidth{\EdgeDblArrowWidth}
    \ChgEdgeArrowLengthCoef{\EdgeDblArrowLengthCoef}}
\newcommand{\EdgeLineSimple}%
   {\EdgeLineDblfalse \RstEdgeArrowWidth \RstEdgeArrowLengthCoef}
\newcommand{\ChgEdgeLabelColor}[1]{\renewcommand{\EdgeLabelCol}{#1}}
\newcommand{\RstEdgeLabelColor}{\ChgEdgeLabelColor{\EdgeLabelColor}}
\newcommand{\SetEdgeLabelColor}[1]%
   {\renewcommand{\EdgeLabelColor}{#1}\RstEdgeLabelColor}
\newcommand{\ChgEdgeLabelScale}[1]{\renewcommand{\EdgeLabelSca}{#1}}
\newcommand{\RstEdgeLabelScale}{\ChgEdgeLabelScale{1}}
\newcommand{\SetEdgeLabelScale}[1]%
   {\renewcommand{\EdgeLabelScale}{#1}\RstEdgeLabelScale}
\newcommand{\FixDimEdge}[4]{%
    \renewcommand{\DimEdgeLineStyle}{#1}%
    \renewcommand{\DimEdgeLineCoef}{#2}%
    \renewcommand{\DimEdgeLineColor}{#3}%
    \renewcommand{\DimEdgeLabelColor}{#4}}%
\newcommand{\ChgEdgeArrowStyle}[1]{\renewcommand{\EdgeArrowSty}{#1}}
\newcommand{\RstEdgeArrowStyle}{\ChgEdgeArrowStyle{\EdgeArrowStyle}}
\newcommand{\SetEdgeArrowStyle}[1]%
   {\renewcommand{\EdgeArrowStyle}{#1}\RstEdgeArrowStyle}
\newcommand{\ChgEdgeArrowWidth}[1]%
   {\setlength{\EdgeArrowSZDim}{#1}} 
\newcommand{\RstEdgeArrowWidth}{\ChgEdgeArrowWidth{\EdgeArrowWidth}}
\newcommand{\SetEdgeArrowWidth}[1]%
   {\setlength{\EdgeArrowWidth}{#1} \RstEdgeArrowWidth}
\newcommand{\ChgEdgeArrowLengthCoef}[1]{\renewcommand{\EdgeArrowSZNum}{#1}}
\newcommand{\RstEdgeArrowLengthCoef}{\ChgEdgeArrowLengthCoef{\EdgeArrowLengthCoef}}
\newcommand{\SetEdgeArrowLengthCoef}[1]%
   {\renewcommand{\EdgeArrowLengthCoef}{#1}\RstEdgeArrowLengthCoef}
\newcommand{\ChgEdgeArrowInsetCoef}[1]{\renewcommand{\EdgeArrowIns}{#1}}
\newcommand{\RstEdgeArrowInsetCoef}{\ChgEdgeArrowInsetCoef{\EdgeArrowInset}}
\newcommand{\SetEdgeArrowInsetCoef}[1]%
   {\renewcommand{\EdgeArrowInset}{#1}\RstEdgeArrowInsetCoef}
\newcommand{\ReverseArrow}%
   {\ChgEdgeArrowStyle{\EdgeRevArrowStyle}%
    \renewcommand{\EdgeLabelPos}{\EdgeLabelRevPosit}%
    \renewcommand{\ArcLabelPos}{\ArcLabelRevPosit}%
    \renewcommand{\LArcLabelPos}{\LArcLabelRevPosit}%
    \renewcommand{\LoopLabelPos}{\LoopLabelRevPosit}%
    \renewcommand{\CLoopLabelPos}{\CLoopLabelRevPosit}%
    \renewcommand{\InitStateLabelPos}{\InitStateLabelRevPosit}%
    \renewcommand{\FinalStateLabelPos}{\FinalStateLabelRevPosit}}
\newcommand{\StraightArrow}%
   {\ChgEdgeArrowStyle{\EdgeArrowStyle}%
    \renewcommand{\EdgeLabelPos}{\EdgeLabelPosit}%
    \renewcommand{\ArcLabelPos}{\ArcLabelPosit}%
    \renewcommand{\LArcLabelPos}{\LArcLabelPosit}%
    \renewcommand{\LoopLabelPos}{\LoopLabelPosit}%
    \renewcommand{\CLoopLabelPos}{\CLoopLabelPosit}%
    \renewcommand{\InitStateLabelPos}{\InitStateLabelPosit}%
    \renewcommand{\FinalStateLabelPos}{\FinalStateLabelPosit}}
\newcommand{\FixEdgeLineDouble}[2]{%
    \renewcommand{\EdgeLineDblCoef}{#1}%
    \renewcommand{\EdgeLineDblSep}{#2}}
\newcommand{\FixEdgeBorder}[2]{%
    \renewcommand{\EdgeLineBorderCoef}{#1}%
    \renewcommand{\EdgeLineBorderColor}{#2}}
\newcommand{\EdgeBorder}%
  {\setlength{\EdgeLineBord}{\EdgeLineBorderCoef\EdgeLineWid}}

\newcommand{\ChgZZLineWidth}[1]{\renewcommand{\ZZLineWid}{#1}}
\newcommand{\RstZZLineWidth}{\ChgZZLineWidth{\ZZLineWidth}}
\newcommand{\SetZZLineWidth}[1]%
   {\renewcommand{\ZZLineWidth}{#1}\RstZZLineWidth}
\newcommand{\VaucEdgeLabel}[1]{%
        \textcolor{\EdgeLabelCol}{\scalebox{\EdgeLabelSca}{\scalebox{\EdgeLabelScale}{$ #1 $}}}}
\newcommand{\RstEdge}{%
   \RstEdgeOffset\RstArcAngle\RstLArcAngle%
   \RstArcCurvature\RstLArcCurvature%
   \RstArcOffset\RstLoopOffset\RstLoopSize%
   \RstEdgeLineColor\RstEdgeLineStyle\RstEdgeLineWidth\EdgeLineSimple%
   \StraightArrow
   \RstEdgeLabelScale\RstEdgeLabelColor}

\newcommand{\Initial}[2][\InitialDir]{\EdgeStyle\ncline{#2#1}{#2}}
\newcommand{\Final}[2][\FinalDir]{\EdgeStyle\ncline{#2}{#2#1}}

\newcommand{\InitialL}[4][{\InitStateLabelPos}]%
   {\EdgeStyle\ncline{#3#2}{#3}\naput[npos=#1]{\VaucEdgeLabel{#4}}}
\newcommand{\InitialR}[4][{\InitStateLabelPos}]%
   {\EdgeStyle\ncline{#3#2}{#3}\nbput[npos=#1]{\VaucEdgeLabel{#4}}}
\newcommand{\FinalL}[4][{\FinalStateLabelPos}]%
   {\EdgeStyle\ncline{#3}{#3#2}\naput[npos=#1]{\VaucEdgeLabel{#4}}}
\newcommand{\FinalR}[4][{\FinalStateLabelPos}]%
   {\EdgeStyle\ncline{#3}{#3#2}\nbput[npos=#1]{\VaucEdgeLabel{#4}}}
\newcommand{\EdgeL}[4][{\EdgeLabelPos}]%
   {\EdgeStyle \ncline{#2}{#3} \naput[npos=#1]{\VaucEdgeLabel{#4}}}
\newcommand{\EdgeR}[4][{\EdgeLabelPos}]%
   {\EdgeStyle \ncline{#2}{#3} \nbput[npos=#1]{\VaucEdgeLabel{#4}}}
\newcommand{\ArcL}[4][{\ArcLabelPos}]%
   {\EdgeStyle \psset{style=VaucArcL}%
    \ncarc{#2}{#3} \naput[npos=#1]{\VaucEdgeLabel{#4}}}
\newcommand{\ArcR}[4][{\ArcLabelPos}]%
   {\EdgeStyle \psset{style=VaucArcR}%
    \ncarc{#2}{#3} \nbput[npos=#1]{\VaucEdgeLabel{#4}}}
\newcommand{\LArcL}[4][{\LArcLabelPos}]%
   {\EdgeStyle \psset{style=VaucLArcL}%
    \ncarc{#2}{#3} \naput[npos=#1]{\VaucEdgeLabel{#4}}}
\newcommand{\LArcR}[4][{\LArcLabelPos}]%
   {\EdgeStyle \psset{style=VaucLArcR}%
    \ncarc{#2}{#3} \nbput[npos=#1]{\VaucEdgeLabel{#4}}}
\newcounter{anglea}\newcounter{angleb}
\newcommand{\LoopXR}[7]%
   {{\setcounter{anglea}{#2-#4}}%
    {\setcounter{angleb}{#2+#4}}%
    {\EdgeStyle \psset{angleA=\theanglea,angleB=\theangleb,offset=#5,ncurv=#6}%
    \nccurve{#3}{#3} \nbput[npos=#1]{\VaucEdgeLabel{#7}}}}
\newcommand{\LoopXL}[7]%
   {{\setcounter{anglea}{#2+#4}}%
    {\setcounter{angleb}{#2-#4}}%
    {\EdgeStyle \psset{angleA=\theanglea,angleB=\theangleb,offset=-#5,ncurv=#6}%
    \nccurve{#3}{#3} \naput[npos=#1]{\VaucEdgeLabel{#7}}}}
\newcommand{\LoopR}[4][{\LoopLabelPos}]%
   {\LoopXR{#1}{#2}{#3}{\LoopAng}{\LoopOff}{\LoopSi}{#4}}
\newcommand{\LoopL}[4][{\LoopLabelPos}]%
   {\LoopXL{#1}{#2}{#3}{\LoopAng}{\LoopOff}{\LoopSi}{#4}}
\newcommand{\CLoopR}[4][{\CLoopLabelPos}]%
   {\LoopXR{#1}{#2}{#3}{\CLoopAng}{\LoopOff}{\LoopSi}{#4}}
\newcommand{\CLoopL}[4][{\CLoopLabelPos}]%
   {\LoopXL{#1}{#2}{#3}{\CLoopAng}{\LoopOff}{\LoopSi}{#4}}
\newcommand{\LoopVarR}[4][{\LoopLabelPos}]%
   {\LoopXR{#1}{#2}{#3}{\LoopVarAng}{\LoopVarOff}{\LoopVarSi}{#4}}
\newcommand{\LoopVarL}[4][{\LoopLabelPos}]%
   {\LoopXL{#1}{#2}{#3}{\LoopVarAng}{\LoopVarOff}{\LoopVarSi}{#4}}

\newcommand{\LoopN}[3][{\LoopLabelPos}]{\LoopL[#1]{90}{#2}{#3}}
\newcommand{\LoopS}[3][{\LoopLabelPos}]{\LoopR[#1]{-90}{#2}{#3}}

\newcommand{\ZZEdge}[2]%
   {\EdgeStyle\ZigzagStyle\nczigzag{#1}{#2}}%
\newcommand{\ZZEdgeL}[4][{\EdgeLabelRevPosit}]%
   {\EdgeStyle\ZigzagStyle\nczigzag{#2}{#3}%
    \naput[npos=#1]{\VaucEdgeLabel{#4}}}
\newcommand{\ZZEdgeR}[4][{\EdgeLabelRevPosit}]%
   {\EdgeStyle\ZigzagStyle\nczigzag{#2}{#3}%
    \nbput[npos=#1]{\VaucEdgeLabel{#4}}}

\newcommand{\VArcL}[5][{\ArcLabelPos}]%
   {\EdgeStyle \psset{style=VaucLArcL}%
    \ncarc[#2]{#3}{#4} \naput[npos=#1]{\VaucEdgeLabel{#5}}}
\newcommand{\VArcR}[5][{\ArcLabelPos}]%
   {\EdgeStyle \psset{style=VaucLArcR}%
    \ncarc[#2]{#3}{#4} \nbput[npos=#1]{\VaucEdgeLabel{#5}}}
\newcommand{\VCurveL}[5][{\ArcLabelPos}]%
   {\EdgeStyle \psset{angleA=0,angleB=180,ncurv=1}%
    \nccurve[#2]{#3}{#4} \naput[npos=#1]{\VaucEdgeLabel{#5}}}
\newcommand{\VCurveR}[5][{\ArcLabelPos}]%
   {\EdgeStyle \psset{angleA=0,angleB=0,ncurv=1}%
    \nccurve[#2]{#3}{#4} \nbput[npos=#1]{\VaucEdgeLabel{#5}}}
\newcommand{\LabelL}[2][{\EdgeLabelPos}]%
   {\naput[npos=#1]{\VaucEdgeLabel{#2}}}
\newcommand{\LabelR}[2][{\EdgeLabelPos}]%
   {\nbput[npos=#1]{\VaucEdgeLabel{#2}}}
\SetStateLabelColor{black}              
\SetStateLabelScale{1.7}                
\SetStateLineStyle{solid}               
\SetStateLineWidth{1.8pt}               
\SetStateLineColor{black}               
\SetStateFillStatus{solid}              
\SetStateFillColor{white}               
\FixDimState{solid}{gray}{1}{gray}{white}  
\FixStateLineDouble{0.6}{0.4}               
\SetEdgeLabelColor{black}               
\SetEdgeLabelScale{1.7}                 
\SetEdgeLineStyle{solid}                
\SetEdgeLineWidth{1pt}                  
\SetEdgeLineColor{black}                
\SetArcAngle{15}                        
\SetLArcAngle{30}                       
\SetArcCurvature{0.8}                   
\SetEdgeOffset{0pt}                     
\SetArcOffset{1pt}                      
\SetLoopOffset{0pt}                     
\renewcommand{\ForthBackEdgeOffset}{5}  
\FixDimEdge{solid}{1.2}{gray}{gray}     
\FixEdgeBorder{2}{white}                
\FixEdgeLineDouble{0.4}{0.8}            
\renewcommand{\ZZShape}{0.5}            
\SetZZLineWidth{1.7}                    
\renewcommand{\TransLabelZZCoef}{0.6}   
\renewcommand{\MediumScale}{0.6}        
\renewcommand{\VSStateLineCoef}{.6}             
\renewcommand{\ArrowOnMediumState}{1.5}         
\renewcommand{\ArrowOnSmallState}{1.7}          
\renewcommand{\ArrowOnLargeState}{1.3}          
\renewcommand{\ArrowOnVerySmallState}{5}        
\renewcommand{\LoopOnMediumState}{7}            
\renewcommand{\LoopOnSmallState}{9.6}           
\renewcommand{\LoopOnLargeState}{5.8}           
\renewcommand{\CLoopOnMediumState}{8}           
\renewcommand{\CLoopOnSmallState}{12}           
\renewcommand{\CLoopOnLargeState}{6}            
\renewcommand{\EdgeLabelPosit}{.45}   
\renewcommand{\EdgeLabelRevPosit}{.55}
\renewcommand{\ArcLabelPosit}{.40}

\renewcommand{\LArcLabelPosit}{.40}

\renewcommand{\LoopLabelPosit}{.25}

\renewcommand{\CLoopLabelPosit}{.25}

\renewcommand{\InitStateLabelPosit}{.10}

\renewcommand{\FinalStateLabelPosit}{.90}

\SetEdgeArrowWidth{5pt}         
\SetEdgeArrowLengthCoef{1.4}            
\renewcommand{\EdgeDblArrowLengthCoef}{1.7}     
\SetEdgeArrowInsetCoef{0.1}                     
\SetEdgeArrowStyle{->}                          
\renewcommand{\EdgeRevArrowStyle}{<-}           
\renewcommand{\StateDimen}{outer}               
\renewcommand{\StateDblDimen}{middle}           
\SetVCDirectory{}        
\FixVCScale{.6}
\HideFrame
\HideGrid
\MediumPicture
\HideName
\RigidLabel
\FullState                              
\MediumState

 \newtheorem{thrm}{Theorem}[section]
 \newtheorem{prpstn}[thrm]{Proposition}

\newtheorem{crllr}[thrm]{Corollary}

\newcommand{\mrm}[1]{\text{\rm #1}}
\newcommand{\supp}[1]{\text{Supp }#1}

\def\cA{{\mathcal A}}
\def\cU{{\mathcal U}}

\def\cT{{\mathcal T}}

\def\cD{{\mathcal D}}

\def\cB{{\mathcal B}}

\def\cP{{\mathcal P}}
\def\cS{{\mathcal S}}

\newcommand{\N} {\ensuremath{\mathbb{N}}}
\newcommand{\Z} {\ensuremath{\mathbb{Z}}}

\newcommand{\R} {\ensuremath{\mathbb{R}}}
\newcommand{\K} {\ensuremath{\mathbb{K}}}
\newcommand{\B} {\ensuremath{\mathbb{B}}}
\newcommand{\Nmin} {\N_{\min}}

\newcommand{\Zmin} {\Z_{\min}}
\newcommand{\Zmax} {\Z_{\max}}

\newcommand{\Rmin} {\R_{\min}}
\newcommand{\Rmax} {\R_{\max}}

\newcommand{\1}{\mathbb{1}}

\newcommand{\fleche}[2]{\mathchoice
         {\xrightarrow{#1\mid #2}}
         {\xrightarrow{\smash{\lower1pt\hbox{$\scriptstyle #1$}}}}
         {\text{Erreur}}
         {\text{Erreur}}}

\newcommand{\wght}[1]{\textit{weight}\left(#1\right)}

\newcommand{\coef}[2]{\langle #1, #2\rangle}

\def\Rat{\text{Rat}}

\def\ab{\Sigma}

\def\A{\mathcal{A}}

\begin{document}
\title{Series which are both max-plus and min-plus rational are
  unambiguous}
\author{Sylvain Lombardy and Jean Mairesse\thanks{%
LIAFA (UMR 7089), CNRS - Universit\'e Paris 7,
2 place Jussieu, 75251 Paris cedex 05, France,
e-mail: {\small{\tt \{lombardy,mairesse\}@liafa.jussieu.fr}}}}

\maketitle

\begin{abstract}
Consider partial maps $\Sigma^* \longrightarrow \R$ with a rational
domain. We show that two families of such series are actually the
same: the unambiguous rational series on the one hand, and 
the max-plus and min-plus rational series on the other hand.
The decidability of equality was known to hold in both families with
different proofs, so the above unifies the picture. 
We give an effective procedure to build an unambiguous automaton from
a max-plus automaton and a min-plus one that recognize the same series.
\end{abstract}

\section{Introduction}

A max-plus automaton is an automaton with multiplicities in the
semiring $\Rmax= (\R\cup\{-\infty\},\max,+)$. Roughly, the transitions
of the automaton have a {\em label} in a finite alphabet $\Sigma$ and a {\em
  weight} in the semiring. The weight of a word $w$ in $\ab^*$ is the maximum over all
successful paths of label $w$ of the sum of the weights along the path. 
The series {\em recognized} by the
automaton $\cT$ is the resulting map $S(\cT): \Sigma^* \rightarrow
\Rmax$. The set of series recognized by a max-plus automaton 
is denoted by $\Rmax\Rat(\Sigma^*)$. 

These automata, or the variants obtained by considering the min-plus
semi\-ring $\Rmin= (\R\cup\{+\infty\},\min,+)$ or subsemirings such as 
$\Zmax$ or the {\em tropical semiring} $\Nmin$, have been studied
under various names: distance automata, cost automata, finance
automata... 
The motivations range from complexity issues in formal language
theory~\cite{simo88}, 
to automatic speech recognition \cite{mohr}, via the modeling of Tetris
heaps \cite{GaMa98b}. 

In Krob~\cite{krob94}, the following question was raised: characterize the
series which are recognized both by a max-plus and a min-plus
automaton. That is, characterize the class $\Rmax\Rat(\Sigma^*)\cap
\Rmin\Rat(\Sigma^*)$. Here, we answer the question by showing that these
series are precisely the unambiguous max-plus (equivalently, min-plus)
series. 
Given a finitely ambiguous max-plus automaton, it is decidable if the
corresponding series is unambiguous~\cite{KLMP04}. On the other hand,
the status of the same problem
starting from an infinitely ambiguous max-plus automaton is unknown.

Apart from an interest in terms of classification, this result 
clarifies the status of the equality problem for 
max-plus series. The equality problem is to determine if ``$S=T$'',
where $S$ and $T$ are series recognized by given max-plus
automata. The equality problem is already undecidable in $\Zmax$ and
for two letters alphabet~\cite{krob}, but it is decidable for
finitely ambiguous automata over $\Rmax$~\cite{HIJi,webe94}. Also, the following result
is proved in \cite{krob94}: if $\cA$ is an automaton over $\Zmax$, and
$\cB$ an automaton over $\Zmin$, then the problem ``$S(\cA)=S(\cB)$'' is
decidable, so the equality problem is decidable 
in $\Zmax\Rat(\Sigma^*)\cap \Zmin\Rat(\Sigma^*)$ (see Proposition \ref{pr-maxmin}). We can now
conclude that the decidability result in \cite{krob94} is a particular
case of the one in \cite{HIJi,webe94}. 

\medskip

The paper is organized as follows. In \S \ref{se-krob}, we extend several
results of \cite{krob94} from $\Zmax$ to $\Rmax$, in particular the
so-called Fatou property. The
results are then used in \S \ref{se-minmax} to obtain the characterization of
$\Rmax\Rat(\Sigma^*)\cap \Rmin\Rat(\Sigma^*)$. 

\medskip

Below, the results on decidability and
complexity should be interpreted under the assumption that two real
numbers can be added or compared in constant time. 

\section{Preliminaries}\label{se-prel}

Let $\mathbb{K}$ be any semiring and denote the neutral element of
the additive, resp. multiplicative, law by $0_\mathbb{K}$,
resp. $1_\mathbb{K}$. 
Let $Q$ be a finite set and $\ab$ a finite alphabet.
A finite linear representation indexed by $Q$ over the alphabet $\ab$
and the semiring $\mathbb{K}$
is a triple $(\alpha,\mu,\beta)$, where
$\alpha$, resp. $\beta$, is a row, resp. columm, vector
of $\K^{Q}$ and $\mu$ is a morphism from $\ab^*$ into $\K^{Q\times Q}$
(for $u=u_1\cdots u_n, u_i\in \Sigma, \ \mu(u)=\mu(u_1)\cdots \mu(u_n)$).
The (formal power) series {\it recognized} by $(\alpha,\mu,\beta)$ is the
series~$S:\ab^* \rightarrow \K$ such that $\coef{S}{w}=\alpha\mu(w)\beta$.
By the Sch\"{u}tzenberger
Theorem, the set of series that can be
recognized by a finite linear representation is precisely the set of
rational series. We denote it by $\mathbb{K}\mrm{Rat}(\ab^*)$.

\medskip

Let $(\alpha,\mu,\beta)$ be a finite linear representation indexed by $Q$
over the semiring $\mathbb{K}$. This representation can be viewed as an automaton
with set of states $Q$:
for every~$(p,q)$ in $Q^2$ and every letter $a$ in
$\ab$, if $\mu(a)\neq 0_\mathbb{K}$, there is a transition from $p$ to $q$
with label $a$ and weight $\mu(a)$. For every $p$ in $Q$,
if $\alpha_p\neq 0_\mathbb{K}$, (resp. $\beta_p\neq 0_\mathbb{K}$),
the state $p$ is initial with weight $\alpha_p$ (resp. terminal
with weight $\beta_p$). In the sequel, we identify the linear representation
with the corresponding automaton.
As usual we transfer the terminology of graph theory to automata, {\it e.g.}
(simple) path or circuit of an automaton. A path which is both starting with an ingoing arc and 
ending with an outgoing arc is called a {\em successful path}.  
The {\em label of a path} is the concatenation of the labels of the
successive arcs (transitions). The {\em weight of a 
  path} is the product (with respect to the multiplicative law of the
semiring) of the weights of the successive 
arcs (including the ingoing and the outgoing arc, need it be).
We denote by $\wght{\pi}$ the weight of the path $\pi$.

Two automata are {\em equivalent} if they recognize the same series. 

\medskip

The {\em support} of a series~$S$ is the set of words~$w$ such that
 $\coef{S}{w}\neq0_\mathbb{K}$. We denote the support of $S$ by $\supp{S}$. The
 characteristic series of a language $L$ is the series 
  $\1_L$ such that $\coef{\1_L}{w}=1_\mathbb{K}$ if $w\in L$, and
  $\coef{\1_L}{w}=0_\mathbb{K}$ otherwise.

\medskip

The {\em max-plus semiring} $\Rmax$ is the semiring formed by the set
$\R\cup\{-\infty\}$ with $\max$ as 
the additive operation and $+$ as the multiplicative operation.
In the sequel, we sometimes denote $\max$ and $+$ respectively by $\oplus$ and
$\otimes$; the neutral elements for these operations are respectively
$-\infty$ and $0$. This semiring
is naturally ordered by the usual order on $\R$ extended by: $\forall
a, -\infty \leq a$. 
The {\em min-plus semiring} $\R_{\min}$ is obtained by replacing
$\max$ by $\min$ and $-\infty$ by $+\infty$
in the definition of $\Rmax$.
The subsemirings $\Rmax^{-}$, $\Zmax$, $\Zmax^{-}$, $\Zmin$, \ldots, are defined in the natural
way. 

The subsemiring $\B=\{(-\infty,0), \oplus, \otimes\}$ of $\Rmax$ is
the Boolean semiring. 
There exists a
morphism from $\Rmax$ 
onto $\B$ that maps $-\infty$ onto $-\infty$ and
any other element onto $0$.

\medskip

An automaton over $\Rmax$ is called a {\em max-plus
automaton}, the corresponding series is called a {\em max-plus
(rational) series}.
Let $S$ be a max-plus
  rational series recognized by
$(\alpha,\mu,\beta)$. Then $\supp S$ is the regular language 
recognized
by the Boolean automaton obtained from $(\alpha,\mu,\beta)$ by
applying to each coefficient the canonical morphism from $\Rmax$ onto
$\B$. 

\medskip

An automaton is {\em unambiguous} if, for every word $w$,
there is at most one successful path labeled by $w$.
An automaton is {\em 1-valued} if, for every word $w$,
all the successful paths labeled by $w$ have the same weight.

In $\Rmax$, a triple $(\alpha,\mu,\nu)$ is unambiguous if, for every word $w$,
\begin{align*}
a) & \ \text{there exists at most one } i,\  (\alpha\mu(w))_i+\beta_i\neq -\infty\\
b) & \ \forall a\in\Sigma,\ \forall j, \ \text{there exists at most
  one }
i,\ (\alpha\mu(w))_i+\mu(a)_{ij}\neq -\infty \:.
\end{align*}
In $\Rmax$, a triple $(\alpha,\mu,\nu)$ is 1-valued if,
for every word $w$,
\begin{align*}
a) & \ \exists x_w,\ \forall i,\  (\alpha\mu(w))_i+\beta_i\in\{-\infty,x_w\}\\
b) & \ \forall a\in\Sigma,\ \forall j,\ \exists x,\ \forall i,\
(\alpha\mu(w))_i+\mu(a)_{ij}\in\{-\infty,x\}\:. 
\end{align*}
Analogous definitions hold for triples over $\Rmin$. 
A max-plus, resp. min-plus, series is {\em unambiguous} if there exists 
an unambiguous max-plus, resp. min-plus, automaton recognizing it.

\medskip

The operations on matrices over $\Rmax$ are defined classically
with respect to the operations of $\Rmax$, e.g.:
$(M\otimes N)_{ij}=\bigoplus_k M_{ik}\otimes
M_{kj}=\max_k(M_{ik}+M_{kj})$. We usually write $AB$ for $A\otimes B$. 
Given $u=(u_1,\dots , u_n) \in \Rmax^n$ and $\lambda \in \Rmax$, set $\lambda u
= (\lambda \otimes u_1,\dots , \lambda\otimes u_n)= (\lambda +
u_1,\dots , \lambda + u_n)$. 

\medskip

Consider a matrix $A\in \Rmax^{Q\times Q}$. The matrix $A$ is {\em
  irreducible} if the graph of $A$ (nodes $Q$, $i\rightarrow j$ if
  $A_{ij}\neq -\infty$) is strongly connected. A  scalar $\lambda\in \Rmax$
and a column vector $u \in \Rmax^Q
\setminus (-\infty,\dots, -\infty)$ such that 
\[
A u = \lambda u = (\lambda+u_i)_{i\in Q}\:,
\]
are called respectively an {\em eigenvalue} and an {\em
  eigenvector} of $A$. 
The number of eigenvalues is at least one and at most
  $|Q|$, and it is exactly one if $A$ is irreducible. 
The max-plus spectral theory is the study of
  these eigenvalues and eigenvectors. 
In the sequel, we only need the result in Theorem \ref{th-spectral}.
For a more complete picture, as
  well as proofs and bibliographic references, see for instance \cite{BCOQ}. 

\begin{thrm}[Max-plus spectral theory]\label{th-spectral}
Consider $A\in \Rmax^{Q\times Q}$. Let  $\rho(A)$ be the maximal eigenvalue
of $A$. We have:
\[
\rho(A) = \max_{k \leq |Q|} \quad \max_{i_1,\dots, i_{k-1}\in Q} \frac{A_{i_1i_2}+
   A_{i_2i_3} + \cdots A_{i_{k-1}i_1}}{k}= \max_{k \leq |Q|} \ 
   \max_{i\in Q} \ \frac{A^k_{ii}}{k}\:.
\]
In words, $\rho(A)$ is the maximal mean weight of a simple circuit of
(the graph of) $A$. 
\end{thrm}

\section{Some decidability results}\label{se-krob}

In this section, we reconsider the various results proved by
Krob~\cite{krob94} for series in $\Zmax$ and we extend
them to $\Rmax$. The proofs are different since they use
the max-plus spectral theory. The results are then used in \S
\ref{se-minmax}. Obviously, analogous results hold for $\Rmin$. 

\medskip

The decidability part of Proposition \ref{pr-ineq} is
given in \cite[Corollary 4.3]{krob94} for
series in $\Zmax\mrm{Rat}(\ab^*)$.
The proof in \cite{krob94} is different and relies
on the fact that $\Zmax\mrm{Rat}(\ab^*)$ is a constructive Fatou
extension of $\Zmax^{-}\mrm{Rat}(\ab^*)$. We prove a generalization of
this last 
result for  $\Rmax\mrm{Rat}(\ab^*)$ in Proposition
\ref{pr-fatou} below. Using Proposition \ref{pr-fatou}, we can then 
recover the decidability in Proposition \ref{pr-ineq} in the same way
as in \cite{krob94}. 
Observe however that the proof of Proposition \ref{pr-ineq} 
given below provides a polynomial
procedure. 

In contrast with Proposition \ref{pr-ineq}, the problem ``$\forall w\in \ab^*,  
\coef{S}{w} \geq 0$'' is undecidable even for $S \in \Zmax\Rat(\ab^*)$, see
\cite{krob}. 

\begin{prpstn}
\label{pr-ineq}
Consider the following problem:
\begin{tabbing} 
de la marge \= \kill
\> {\bf Instance:} \hspace*{0.4cm} \= $S\in \Rmax\mrm{Rat}(\ab^*)$ \\
\>  {\bf Problem:} \> $\forall w\in \ab^*,  \
\coef{S}{w} \leq 0$ \:.
\end{tabbing}
This problem can be decided with an algorithm of polynomial time
complexity in the size of an automaton recognizing $S$. 
\end{prpstn}

\begin{proof}
Let $\A=(\alpha,\mu,\beta)$ be a trim automaton recognizing $S$ with
set of states $Q$. Set 
\[
M=\bigoplus_{a\in \ab} \mu(a)\:.
\]
Let $\rho(M)$ be the maximal eigenvalue of $M$. 
By the Max-plus Spectral Theorem \ref{th-spectral}, there exist
$k\in \N^*$ and $i\in Q$ such that $M^k_{ii}=k\times \rho(M)$. 
It implies that there exists $w\in \ab^k$ such that 
$\mu(w)_{ii}= k\times \rho(M)$. 
Clearly, we have $\mu(w^n)_{ii} \geq n\times k \times \rho(M)$ for all $n\in \N^*$. 
Since the automaton is 
trim, there exist $w_1,w_2\in \ab^*$ such that $\alpha
\mu(w_1)_i > -\infty$ and  $\mu(w_2)\beta_i > -\infty$.  
Assume that $\rho(M)>0$, then 
by choosing $n$ large enough, we get the following contradiction
\[
\coef{S}{w_1w^nw_2} \geq \alpha
\mu(w_1)_i + \mu(w^n)_{ii} + \mu(w_2)\beta_i > 0 \:.
\]
Assume now that $\rho(M)\leq 0$. By the Max-plus Spectral Theorem
\ref{th-spectral}, 
it implies that all the circuits in the automaton have a weight which
is non-positive. 
Assume that there exists a word $w$ such
that $\coef{S}{w}>0$. Let $\pi$ be a successful path of label $w$ and 
maximal weight in the automaton. If $\pi$ contains a circuit, then 
the path $\pi'$ obtained by removing the circuit is still a successful
path. In particular, if $w'$ is the label of $\pi'$,
we have $\coef{S}{w'} \geq \coef{S}{w}>0$. 
So we can choose, without loss of generality, a word $w$ such that
$\coef{S}{w}>0$ and $|w|<|Q|$. 
Now notice that we have for all $k\in \N$,
\[
\left( \exists u \in \ab^k, \coef{S}{u}>0 \right) \ \iff \ 
\alpha M^k \beta > 0 \:.
\]
Summarizing the results obtained so far, we get
\begin{equation}\label{eq-crit}
\left( \forall u \in \ab^*, \coef{S}{u} \leq 0 \right) \ \iff \ 
\left( \rho(M) \leq 0 \right) \wedge ( \ \forall k \in \{0, \dots ,
|Q|-1\}, \alpha M^k \beta \leq 0 \ )\:,
\end{equation}
where $M^0$ is the identity matrix defined by: $\forall i, M^0_{ii}=0,
\ \forall i\neq j, M^0_{ij}=-\infty$. 

\paragraph{Complexity.}
Computing the matrix $M$ has a time complexity $O(|\ab | |Q|^2)$. 
Computing $\rho(M)$ can be done using Karp algorithm \cite[Theorem
2.19]{BCOQ} in 
time $O(|Q|^3)$. Computing $\alpha M^k \beta$ for all $k\in \{0, \dots ,
|Q|-1\}$ requires also a time complexity $O(|Q|^3)$. 
\end{proof}

Proposition \ref{pr-fatou} is proved for series in
$\Zmax\mrm{Rat}(\ab^*)$ in \cite[Proposition 4.2]{krob94}. It is not obvious to
extend the approach of \cite{krob94} to series in
$\Rmax\mrm{Rat}(\ab^*)$. We propose a quite different proof. 

\begin{prpstn}[Fatou property]\label{pr-fatou}
Consider a series $S$ in $\Rmax\mrm{Rat}(\ab^*)$. Then we have
\[
S: \ab^*\longrightarrow \Rmax^- \ \implies \ S\in \Rmax^-\mrm{Rat}(\ab^*)\:.
\]
Furthermore, given an automaton $\A$ over $\Rmax$ recognizing
$S$, one can effectively compute an automaton $\A^-$ over $\Rmax^-$
recognizing $S$. The procedure to get $\A^-$ from $\A$ has a
polynomial time complexity in the size of $\A$. 
\end{prpstn}

\begin{proof}
Let $(\alpha,\mu,\beta)$ be a trim triple recognizing $S$ with set of
states $\{1,\dots, n\}$. 
Define the matrix $M=\bigoplus_{a\in \ab} \mu(a)$. Since $S:
\ab^*\longrightarrow \Rmax^-$, it follows from (\ref{eq-crit}) that
$\rho(M)\leq 0$. In particular any circuit has non-positive weight. It
follows immediately that:
\[
M^*=\bigoplus_{i\in \N} M^i = I\oplus M \oplus M^2\oplus \cdots \oplus
M^{n-1}\:,
\]
where $I$ is the identity matrix of dimension $n\times n$ defined by
$\forall i, I_{ii}=0, \ \forall i\neq j, I_{ij}=-\infty$. 
Since $S:
\ab^*\longrightarrow \Rmax^-$, it follows that $\alpha M^*\beta\leq 0$.
Set $u=M^*\beta$ and define the diagonal matrix (the non-diagonal coefficients
being $-\infty$) $D= \mrm{diag} (u_1,\dots, u_n)$. Define
\[
\widehat{\alpha}=\alpha D, \quad \widehat{\beta}=D^{-1}\beta, \quad
\forall a \in \ab, \
\widehat{\mu}(a)=D^{-1}\mu(a)D \:.
\]
Clearly, the automaton $(\widehat{\alpha}, \widehat{\mu} , \widehat{\beta})$ 
recognizes the  series $S$.  
We have: $\forall i, \ \widehat{\alpha}_i = \alpha_i + (M^*\beta)_i \leq
\alpha M^*\beta \leq 0$; and also: $\forall i, \ \widehat{\beta}_i =
\beta_i - (M^*\beta)_i \leq \beta_i -\beta_i =0$. At last, we have:
$\forall a\in \ab, \forall i$,
\begin{eqnarray*}
\bigoplus_j \widehat{\mu}(a)_{ij} \ \leq \ \bigoplus_j (D^{-1}MD)_{ij} & =
&  \bigoplus_j \bigl[ (D^{-1}M)_{ij}+ (M^*\beta)_j \bigr] \\ 
& = & 
(D^{-1}MM^*\beta)_i \ \leq \ (D^{-1}M^*\beta)_i\ =\ 0 \:,
\end{eqnarray*}
where we have used that $MM^*\leq I\oplus MM^*=M^*$.
Hence the triple $(\widehat{\alpha}, \widehat{\mu} , \widehat{\beta})$
is defined over the semiring $\Rmax^-$. This completes the proof.

\paragraph{Complexity.}
The matrix $M$ is computed in time $O(|\ab|n^2)$. 
Then, computing $u=M^*\beta$ requires $O(n^3)$ operations. Knowing $u$,
computing 
$(\widehat{\alpha}, \widehat{\mu} , \widehat{\beta})$ requires
$O(|\ab|n^3)$ operations. 
\end{proof}

Proposition \ref{pr-eq} is proved for series in
$\Zmax\mrm{Rat}(\ab^*)$ in \cite[Proposition 5.1]{krob94}.
The proof relies on the Fatou property. Since
we have extended this last property to $\Rmax\mrm{Rat}(\ab^*)$, the
proof of Krob carries over unchanged. In the proof below, we present
the arguments in a slightly different way. 

\begin{prpstn}
\label{pr-eq}
The following problem is decidable:
\begin{tabbing} 
de la marge \= \kill
\> {\bf Instance:} \hspace*{0.4cm} \= $S\in \Rmax\mrm{Rat}(\ab^*)$ and
$c\in \R$ \\
\>  {\bf Problem:} \> $\forall w\in \ab^*,  \ \coef{S}{w} = c$ \ \
(i.e. $S=c$)\:.
\end{tabbing}
\end{prpstn} 

\begin{proof}
First of all, it is enough to prove the result for $c=0$. Indeed,
testing if $\coef{S}{w}=c$ is equivalent to testing if
$\coef{S'}{w}=0$ where $S'$ is  
the series defined by $\coef{S'}{u}=\coef{S}{u} - c$. And it is
straightforward to get a triple recognizing $S'$ from a triple
recognizing $S$. 

According to 
Proposition \ref{pr-ineq}, we can decide if $S$ belongs to
$\Rmax^-(\ab^*)$. If not, then we have $S\neq 0$. If $S \in
\Rmax^-(\ab^*)$ then, by Proposition \ref{pr-fatou}, there exists an
effectively computable automaton $(\alpha, \mu, \beta)$ over $\Rmax^-$ recognizing $S$. 
We define an automaton
$(\overline{\alpha}, \overline{\mu}, 
\overline{\beta})$ as follows: 
\[
\forall a\in \ab, \forall i, j, \ \overline{\mu}(a)_{ij}= 
\begin{cases}
0 & \mrm{if } \mu(a)_{ij}= 0 \\
-\infty & \mrm{if }  \mu(a)_{ij} < 0 
\end{cases}\:,
\]
with $\overline{\alpha}$ and $\overline{\beta}$ being defined from
$\alpha$ and $\beta$ in the same
way. 
The important property is that for $w\in \ab^*$, 
\begin{equation}\label{eq-imp}
\coef{S}{w} = 0 \ \iff \ \overline{\alpha} \ \overline{\mu}(w)
\overline{\beta} =0 \:.
\end{equation}
Let us set 
$\overline{\mu}(\ab^*) = \{\overline{\mu}(w), w\in
\ab^*\}$. Obviously, $(\overline{\mu}(\ab^*), \otimes)$ is a submonoid
of the finite monoid $(\B^{n\times n}, \otimes)$. In particular,
$\overline{\mu}(\ab^*)$ is finite and can be effectively
constructed. In view of (\ref{eq-imp}), we have
\begin{equation}\label{eq-univ}
\left(\forall w \in \ab^*,  \coef{S}{w} = 0 \right) \ \iff \ 
\left( \forall A \in \overline{\mu}(\ab^*), \overline{\alpha} A
\overline{\beta} = 0 \right)\:. 
\end{equation}
Since $\overline{\mu}(\ab^*)$ is finite and effectively computable,
the property on the right can be checked algorithmically. 
\end{proof}

\paragraph{Complexity.}
In contrast with Proposition \ref{pr-ineq}, we do not get a polynomial
procedure in Proposition \ref{pr-eq}. 
Deciding if the right-hand side in (\ref{eq-univ}) holds is PSPACE-complete with respect to the
dimension of the triple, see for instance \cite[Theorem 13.14 and
  Exercise 13.25]{HoUl}.  This is known as 
the {\em universality problem}.

\begin{prpstn}
\label{pr-eq-part}
The following problem is decidable:
\begin{tabbing} 
de la marge \= \kill
\> {\bf Instance:} \hspace*{0.4cm} \= $S\in \Rmax\mrm{Rat}(\ab^*)$ and
$c\in \R$ \\
\>  {\bf Problem:} \> $\forall w\in \supp{S},  \ \coef{S}{w} = c$\;.
\end{tabbing}
\end{prpstn}
\begin{proof}
The proof is the same as in Proposition~\ref{pr-eq}.
Instead of deciding the right-hand side of (\ref{eq-univ}), it must be decided whether
$(\alpha,\mu,\beta)$ and $(\overline\alpha,\overline\mu,\overline\beta)$
have the same support.
\end{proof}

\paragraph{Complexity.}
The complexity of this problem is PSPACE-complete. 
Indeed, $(\overline\alpha,\overline\mu,\overline\beta)$ is obtained
from $(\alpha,\mu,\beta)$ by deleting some transitions. 
And deciding whether the language accepted by a non-deterministic automaton
remains the same after the deletion of some transitions is
PSPACE-complete. We briefly explain why. 
First, the equivalence problem for non-deterministic Boolean automata
is PSPACE-complete~\cite{StMe},
and thus our problem is in PSPACE.
Next, let $\cA$ be a non-deterministic automaton and
let $\cA'$ be the automaton obtained from $\cA$
by adding a state, initial and terminal, with loops labelled by evey letter.
Deciding whether $\cA'$ is equivalent to $\cA$ is equivalent
to deciding whether $\cA$ accepts every word (universality problem),
which is PSPACE-complete.
Thus our problem is PSPACE-hard.

\medskip

Proposition \ref{pr-maxmin} is proved for series in
$\Zmax\mrm{Rat}(\ab^*)$ and $\Zmin\mrm{Rat}(\ab^*)$ in
\cite[Proposition 5.3]{krob94}. As discussed in the Introduction, a
consequence of Proposition \ref{pr-maxmin} is that the equality problem is decidable in
$\Rmax\mrm{Rat}(\ab^*) \cap \Rmin\mrm{Rat}(\ab^*)$. Quoting
\cite{krob94}: ``the problem remains to characterize (such)
series''. This is done in \S \ref{se-minmax}. 

\begin{prpstn}
\label{pr-maxmin}
The following problem is decidable:
\begin{tabbing} 
de la marge \= \kill
\> {\bf Instance:} \hspace*{0.4cm} \= $S\in \Rmax\mrm{Rat}(\ab^*),\
T\in \Rmin\mrm{Rat}(\ab^*)$  \\ 
\>  {\bf Problem:} \> $S=T$\:.
\end{tabbing}
The above equality should be interpreted as: $\supp{S}=\supp{T}$ and
$\forall w\in \supp{S}, \coef{S}{w}=\coef{T}{w}$. 
\end{prpstn} 
\begin{proof}
Define the series $-T$ with coefficients in $\R\cup\{+\infty\}$ by
$\coef{-T}{w}=-\coef{T}{w}$ for all $w$. Clearly $-T \in
\Rmax\Rat(\ab^*)$. The above problem is equivalent to:
\[
(a)\ \supp{S}=\supp{T}\quad\text{and}\quad 
(b)\ \forall w\in\supp{S},\
\coef{S-T}{w}=0\:. 
\] 
Point (a) is the problem of equivalence of rational languages and is thus
decidable. 
The series $S-T$ is the Hadamard max-plus product of $S$ and $-T$;
it is recognized by the tensor product of triples recognizing $S$ and $-T$:

Let $(\alpha,\mu,\nu)$ (resp. $(\alpha',\mu',\nu')$) be a trim triple
recognizing $S$ (resp. $-T$) with set of states $Q=\{1,\ldots,n\}$
(resp. $Q'=\{1,\ldots,m\}$). 
Let $(\iota,\pi,\tau)$ be the triple defined on $Q\times Q'$ by:
\begin{equation*}
\iota_{p,q}=\alpha_p+\alpha'_q\qquad
\tau_{p,q}=\nu_p+\nu'_q \qquad
\pi(a)_{(p,q)(r,s)}=\mu(a)_{pr}+\mu'(a)_{qs}\:.
\end{equation*}
By Proposition~\ref{pr-eq-part}, (b) is decidable.
\end{proof}

Using the same proof, one also shows that ``$S\leq T$'' is
decidable. On the other hand, ``$S\geq T$'' is already undecidable for
$S\in \Zmax\Rat(\ab^*)$ and $T\equiv 0$, see \cite{krob}. 

\section{Max-plus and min-plus rational implies
  unambiguous}\label{se-minmax}

To prove that a series recognized by a max-plus and a min-plus
automaton is also recognized by an unambiguous 
automaton, we use an intermediate step which is to prove that it is
recognized by a 1-valued automaton. 

\medskip

Recall that the notion of 1-valuedness of a max-plus automaton has
been defined in \S \ref{se-prel}. 
This notion clearly extends to any automaton with multiplicities over
an idempotent semiring, in particular to a transducer. 
A {\em transducer} $\cT$ is an automaton over the semiring
$\B\Rat(B^*)$. 
The transducer $\cT$ is {\em 1-valued} (or {\em functional}) if 
$|\supp \coef{S(\cT)}{w}| \leq 1$ for all $w$. 
Next result is classical and due to Eilenberg \cite{eile} and
Sch\"utzenberger \cite{schu76}, see \cite[Chapter IV.4]{bers79}: 
a 1-valued transducer can be effectively transformed into an
equivalent unambiguous one. 
The proof of Eilenberg and Sch\"utzenberger easily extends 
to a 1-valued automaton with multiplicities in an idempotent
semiring. 
Here we give a different and simple proof of the same result. 
The argument is basically the same one as in
\cite[Section 4]{KLMP04}.  

\begin{prpstn}\label{pr-weber}
For any max-plus or min-plus 1-valued automaton, there exists 
an unambiguous automaton which recognizes the same series.
\end{prpstn}

\begin{proof}
Let~$\cA=(\alpha,\mu,\nu)$ be a 1-valued automaton and $\cA'$ the underlying
Boolean automaton.
 Let $\cD=(\beta,\delta,\gamma)$ be the determinized automaton of
 $\cA'$ obtained by the subset construction. Let
 $\cS=(\iota,\pi,\tau)$ be the tensor product of $\cA$ and $\cD$:
\begin{gather*}
\iota_{p,q}=\alpha_p + \beta_q, \qquad \tau_{p,q}= \nu_p + \gamma_q,
\qquad \pi(a)_{(p,q)(r,s)} = \mu(a)_{pr} + \delta(a)_{qs} \:.
\end{gather*} 
The automaton $\cS$ is the {\em Sch\"{u}tzenberger covering} of $\cA$,
see \cite{saka98}. 
There is a {\it competition} in $\cS$ if:
\\(a) there exist  $q$, $r$, $s$, $p$, and $p'$ such that $p\neq p'$, 
  $\pi(a)_{(p,q)(r,s)}\neq-\infty$ and $\pi(a)_{(p',q)(r,s)}\neq-\infty$, or
\\(b) there exist $q$, $p$ and $p'$ such that $p\neq p'$,
  $\tau_{p,q}\neq-\infty$ and $\tau_{p',q}\neq-\infty$.

Let $\cU$ be  any automaton obtained from $\cS$ by
removing the minimal number of transitions and/or terminal arrows such
that there is no more competition. 
We claim that $\cU$ is an unambiguous automaton equivalent to $\cA$.
The proof of this claim
can be found in~\cite[Section 4]{KLMP04}.
\end{proof}

As a side remark, the above proof is also clearly valid in any idempotent semiring. 

\medskip

We now have all the ingredients to prove the main result. 

\begin{prpstn}
\label{pr-maxmin2}
Let $S$  be a series in $\Rmax\mrm{Rat}(\ab^*)$. The series $-S$ is in
$\Rmax\mrm{Rat}(\ab^*)$ if and only if the series $S$ is unambiguous. 
\end{prpstn}

\begin{proof}
Let $\cA=(\alpha,\mu,\nu)$, resp. $\cA'=(\alpha',\mu',\nu')$, be a triple
that recognizes $S$, resp. $-S$.
Let  $\cP=(\iota,\pi,\tau)$ be the triple on the semiring $\Rmax \times
\Rmax$ and with set of states $Q\times Q'$ defined by:
\begin{gather*}
\iota_{p,q}=(\alpha_p,\alpha_p+\alpha'_q),\qquad
\tau_{p,q}=(\nu_p,\nu_p+\nu'_q)\\
\pi(a)_{(p,q)(r,s)}=(\mu(a)_{pr},\mu(a)_{pr}+\mu'(a)_{qs})
\end{gather*}
This triple recognizes the series $(S,S-S)=(S, \1_{\supp{S}})$.

For every vector or matrix $x$ with coefficients in $\Rmax^2$,
for $i$ in $\{1,2\}$, we denote $x^{(i)}$, the projection of $x$ with
respect to the $i$-th coordinate.

By Proposition~\ref{pr-fatou} there exists an
automaton $(\iota',\pi',\tau')$ 
equivalent to $(\iota,\pi,\tau)$ and such that
$(\iota'^{(2)},\pi'^{(2)},\tau'^{(2)})$ is over $\Rmax^{-}$ (the first
ccordinate is unmodified: $\iota'^{(1)}= \iota^{(1)}, \pi'^{(1)}=
\pi^{(1)}, \tau'^{(1)} =\tau^{(1)}$).
We define an automaton
$\cB=(\overline{\iota},\overline{\pi},\overline{\tau})$ over the semiring
$\Rmax$ as follows:
\[
\forall a\in \ab, \forall i, j \in Q\times Q', \ \overline{\pi}(a)_{ij}= 
\begin{cases}
\pi^{(1)}(a)_{ij} & \mrm{if } \pi'^{(2)}(a)_{ij}= 0 \\
-\infty & \mrm{if } \pi'^{(2)}(a)_{ij} < 0 
\end{cases}\:,
\]
with $\overline{\iota}$ and $\overline{\tau}$ being defined from
$\iota'$ and $\tau'$ in the same way. 
We claim that $(\overline{\iota},\overline{\pi},\overline{\tau})$ is a
1-valued automaton that recognizes $S$. 

For every word~$w$, every successful path of
$\cB$ labeled by $w$ has a weight equal to the first coordinate $k_1$
of the weight $k$ of a successful path of $\cP$ such that $k_2=0$.
It means that $k_1$ is the weight of a successful path labeled by $w$ in $\cA$
and that $-k_1$ is the weight of a successful path labeled by $w$ in $\cA'$.
Hence, $k_1\leq\coef{S}{w}$ and $-k_1\leq\coef{-S}{w}$,
and so $k_1=\coef{S}{w}$.
Therefore, every successful path of $\cB$
labeled by $w$ has a weight equal to~$\coef{S}{w}$.

Conversely, every word $w$ in $\supp{S}$ labels a successful path in $\cB$.
Indeed, there is a successful path labeled by $w$ with weight $\coef{S}{w}$ in $\cA$,
and a successful path labeled by $w$ with weight $-\coef{S}{w}$ in $\cA'$.
The product of the two paths gives a successful path in
$\cP$ labeled by $w$ with a weight
having a second coordinate equal to $0$, hence, after applying
Proposition~\ref{pr-fatou}, the weight of every transition along this path has
a second coordinate equal to $0$. 

Therefore $\cB$ recognizes the same series as $\cA$. We complete the
proof by applying Proposition \ref{pr-weber}. 
\end{proof}

There is a canonical bijection from $\Rmax$ to $\Rmin$ that consists
in mapping every $x$ different from $-\infty$ onto itself and $-\infty$
onto $+\infty$. This bijection is obviously {\it not} an isomorphism.
With some abuse, we say that a series $S$ of $\Rmax\mrm{Rat}(\ab^*)$ is also
in $\Rmin\mrm{Rat}(\ab^*)$ if its image with respect to the canonical bijection
is in $\Rmin\mrm{Rat}(\ab^*)$.

\begin{crllr}
\label{pr-maxmin3}
A series $S$ is in $\Rmax\mrm{Rat}(\ab^*)\cap\Rmin\mrm{Rat}(\ab^*)$ if and
only if it is unambiguous. Starting from a pair formed by a max-plus
and a min-plus automaton recognizing $S$, one can effectively compute an
unambiguous automaton recognizing $S$. 
\end{crllr}

Observe that given a pair
formed by a max-plus and a min-plus automaton, it can be checked if
they indeed recognize the same series using Proposition
\ref{pr-maxmin}. 

\begin{proof}
Since $S$ is in $\Rmin\mrm{Rat}(\ab^*)$,
$-S$ is in $\Rmax\mrm{Rat}(\ab^*)$ (there is an {\it isomorphism}
from $\Rmax$ onto $\Rmin$ that maps $x$ onto $-x$). This result is
therefore equivalent to Proposition~\ref{pr-maxmin2}. The effective
computation of an unambiguous automaton recognizing $S$ is done in the
proof of Proposition~\ref{pr-maxmin2}. 
\end{proof}

\paragraph{Complexity.} In Corollary \ref{pr-maxmin3}, one gets 
a 1-valued automaton recognizing $S$ of dimension the product of the dimensions
of the max-plus and min-plus automata. This follows directly from the proof of
Proposition \ref{pr-maxmin2}. The time complexity to construct it is
also clearly polynomial. 
On the other hand, the dimension of an unambiguous automaton
recognizing $S$ may be exponential with respect to the dimension of
the 1-valued automaton. 

\section{Examples}

Let $S$ be the series defined by $\coef{S}{w}=\max(|w|_a,|w|_b)$.
This series is obviously max-plus rational. In~\cite{KLMP04}, it is proved 
that $S$ is not unambiguous (section~$3.2$),
and with a different argument that it is not min-plus rational (section~$3.6$).
We know now that both statements are equivalent.

\medskip

We consider now a simple example on which we illustrate the different
steps of our proof. 

\medskip

\begin{figure}[ht]
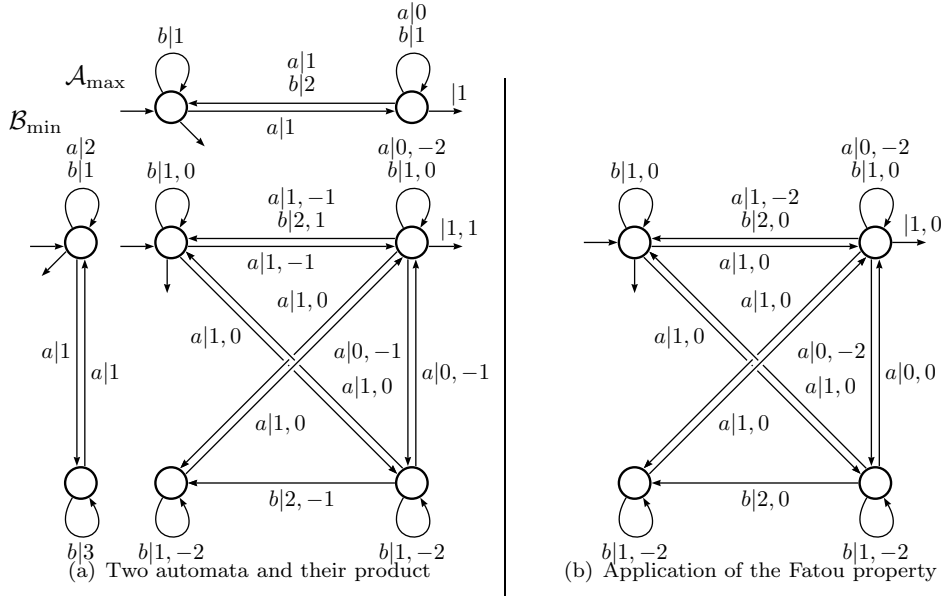

\subfigure[Two automata and their product]{%
\FixVCScale{.5}
\VCDraw[.8]{%
\begin{VCPicture}{(-4.5,-10.5)(10,5)}
\renewcommand{\ForthBackEdgeOffset}{-3}
\renewcommand{\LoopLabelPosit}{.5}
\ForthBackOffset
\VCPut{(-4.5,4)}{\scalebox{2}{${\cal B}_{\min}$}}
\VCPut{(-2.5,5.5)}{\scalebox{2}{${\cal A}_{\max}$}}
\VCPut[-90]{(-3,0)}{%
\State{(0,0)}{A}\State{(8,0)}{B}}
\EdgeR{A}{B}{a|1} \EdgeR{B}{A}{a|1}
\Initial[s]{A}\Final[se]{A}
\LoopN{A}{\StackTwoLabels{a|2}{b|1}}
\LoopS{B}{b|3}
\VCPut[0]{(0,4.5)}{%
\State{(0,0)}{A}\State{(8,0)}{B}}
\Initial{A}\Final[se]{A}\FinalL{e}{B}{|1}
\EdgeR{A}{B}{a|1} \EdgeR{B}{A}{\StackTwoLabels{a|1}{b|2}}
\LoopN{A}{b|1}
\LoopN{B}{\StackTwoLabels{a|0}{b|1}}
\State{(0,0)}{A}\State{(8,0)}{B}\State{(0,-8)}{C}\State{(8,-8)}{D}
\Initial{A}\Final[s]{A}\FinalL{e}{B}{|1,1}
\LoopN{A}{b|1,0}
\LoopN[.5]{B}{\StackTwoLabels{a|0,-2}{b|1,0}}
\LoopS{C}{b|1,-2}
\LoopS{D}{b|1,-2}
\EdgeR{A}{B}{a|1,-1} \EdgeR{B}{A}{\StackTwoLabels{a|1,-1}{b|2,1}}
\EdgeR{B}{D}{a|0,-1} \EdgeR{D}{B}{a|0,-1}
\EdgeBorder
\EdgeR[.3]{A}{D}{a|1,0}
\EdgeR[.3]{B}{C}{a|1,0} 
\EdgeR[.3]{D}{A}{a|1,0}
\EdgeR[.3]{C}{B}{a|1,0}
\RstEdgeOffset
\EdgeL{D}{C}{b|2,-1}
\end{VCPicture}}}
\quad\vline
\subfigure[Application of the Fatou property]{%
\FixVCScale{.5}
\VCDraw[.8]{%
\begin{VCPicture}{(-4,-10.5)(12,5)}
\renewcommand{\LoopLabelPosit}{.5}
\renewcommand{\ForthBackEdgeOffset}{-3}
\State{(0,0)}{A}\State{(8,0)}{B}\State{(0,-8)}{C}\State{(8,-8)}{D}
\Initial{A}\Final[s]{A}\FinalL{e}{B}{|1,0}
\LoopN{A}{b|1,0}
\LoopN{B}{\StackTwoLabels{a|0,-2}{b|1,0}}
\LoopS{C}{b|1,-2}
\LoopS{D}{b|1,-2}
\EdgeL{D}{C}{b|2,0}
\ForthBackOffset
\EdgeR{A}{B}{a|1,0} \EdgeR{B}{A}{\StackTwoLabels{a|1,-2}{b|2,0}}
\EdgeR{B}{D}{a|0,-2} \EdgeR{D}{B}{a|0,0}
\EdgeBorder
\EdgeR[.3]{A}{D}{a|1,0}
\EdgeR[.3]{B}{C}{a|1,0} 
\EdgeR[.3]{D}{A}{a|1,0}
\EdgeR[.3]{C}{B}{a|1,0}
\end{VCPicture}}}
\caption{Getting an unambiguous automaton (I)}\label{fig:ex}
\end{figure}

\begin{figure}[t]
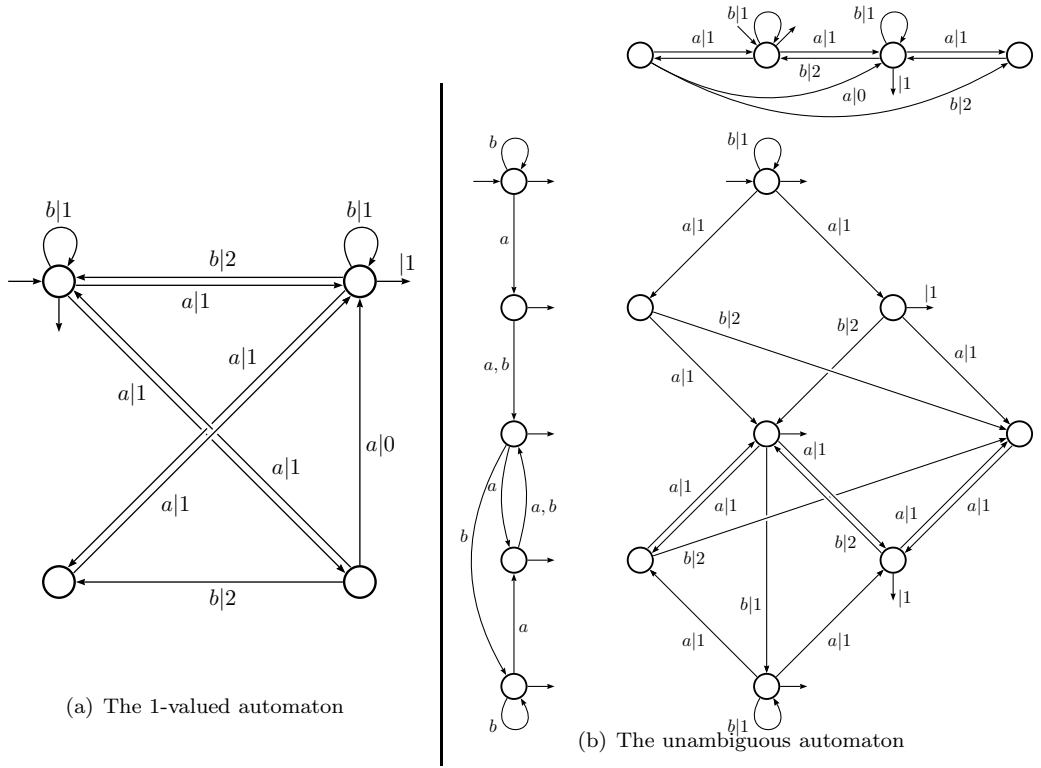

\subfigure[The 1-valued automaton]{%
\FixVCScale{.5}
\VCDraw{%
\begin{VCPicture}{(-2,-11)(10,4)}
\renewcommand{\LoopLabelPosit}{.5}
\renewcommand{\ForthBackEdgeOffset}{-3}
\State{(0,0)}{A}\State{(8,0)}{B}\State{(0,-8)}{C}\State{(8,-8)}{D}
\Initial{A}\Final[s]{A}\FinalL{e}{B}{|1}
\LoopN{A}{b|1}
\LoopN{B}{b|1}
\EdgeL{D}{C}{b|2} \EdgeR{D}{B}{a|0}
\ForthBackOffset
\EdgeR{A}{B}{a|1} \EdgeR{B}{A}{b|2}
\EdgeBorder
\EdgeR[.3]{A}{D}{a|1}
\EdgeR[.3]{B}{C}{a|1} 
\EdgeR[.3]{D}{A}{a|1}
\EdgeR[.3]{C}{B}{a|1}
\end{VCPicture}}}
\vline
\subfigure[The unambiguous automaton]{%
\FixVCScale{.4}
\VCDraw[1.4]{%
\begin{VCPicture}{(-4.5,-1)(9.5,14)}
\State{(-3,12)}{A}
\State{(-3,9)}{B}
\State{(-3,6)}{C}
\State{(-3,3)}{D}
\State{(-3,0)}{E}
\Initial{A}\Final{A}\Final{B}\Final{C}\Final{D}\Final{E}
\LoopN{A}{b}
\EdgeR{A}{B}{a}
\EdgeR{B}{C}{a,b}
\ArcR{C}{D}{a}
\ArcR{D}{C}{a,b}
\LArcR{C}{E}{b}
\EdgeR{E}{D}{a}
\LoopS{E}{b}
\State{(0,15)}{A}
\State{(3,15)}{B}
\State{(6,15)}{C}
\State{(9,15)}{D}
\Initial[nw]{B}\Final[ne]{B}\FinalL[.6]{s}{C}{|1}
\LoopN{B}{b|1}
\LoopN{C}{b|1}
\LArcR[.85]{A}{C}{a|0}
\LArcR[.85]{A}{D}{b|2}
\renewcommand{\ForthBackEdgeOffset}{3}
\ForthBackOffset
\EdgeL{B}{C}{a|1}
\EdgeL[.7]{C}{B}{b|2}
\EdgeL[.5]{A}{B}{a|1}
\EdgeL[]{B}{A}{}
\EdgeL[.5]{C}{D}{a|1}
\EdgeL[]{D}{C}{}
\RstEdgeOffset
\State{(3,12)}{A}
\State{(6,9)}{B}
\State{(0,9)}{C}
\State{(3,6)}{D}
\State{(9,6)}{F}
\State{(3,0)}{G}
\State{(6,3)}{H}
\State{(0,3)}{I}
\Initial{A}\Final{A}\FinalL{e}{B}{|1}\Final{D}\FinalL{s}{H}{|1}\Final{G}
\EdgeBorder
\LoopN{A}{b|1}
\EdgeL{A}{B}{a|1}
\EdgeR{A}{C}{a|1}
\EdgeR[.2]{B}{D}{b|2}
\EdgeL{B}{F}{a|1}
\EdgeR{C}{D}{a|1}
\EdgeL[.2]{C}{F}{b|2}
\EdgeR[.7]{D}{G}{b|1}
\LoopS{G}{b|1}
\EdgeR{G}{H}{a|1}
\EdgeL{G}{I}{a|1}
\EdgeR[.1]{I}{F}{b|2}
\renewcommand{\ForthBackEdgeOffset}{3}
\ForthBackOffset
\EdgeL[.2]{D}{H}{a|1}
\EdgeL[.2]{H}{D}{b|2}
\EdgeL{F}{H}{a|1}
\EdgeL[.2]{H}{F}{a|1}
\EdgeL{D}{I}{a|1}
\EdgeL{I}{D}{a|1}
\end{VCPicture}}}
\caption{Getting an unambiguous automaton (II)}\label{fig:ex2}
\end{figure}

Let $\cA_{\max}$ and $\cB_{\min}$ be the two automata drawn on
Figure~\ref{fig:ex}-(a) (the weights equal to 0 on ingoing or outgoing
arrows have been omitted).
The automaton $\cA_{\max}$ is a max-plus automaton, while
the automaton $\cB_{\min}$ is a min-plus automaton. Their product, performed as
in the proof of Proposition~\ref{pr-maxmin2}, is drawn on the same figure.
The automata are equivalent only if the weight with respect to the second coordinate is
non-positive on every successful path. Hence, we can apply the Fatou property
(Proposition~\ref{pr-fatou}) to get an equivalent automaton on which the
weight ot the second coordinate is non-positive on every arc (transitions, and initial
and final arrows). The result is shown on Figure~\ref{fig:ex}-(b). After
deleting the arcs that have a second coordinate weight different
from~$0$, and remembering only the first coordinate, we get the 
$1$-valued automaton of Figure~\ref{fig:ex2}-(a). As this automaton has the
same support as $\cA_{\max}$ and $\cB_{\min}$, we can conclude that
$\cA_{\max}$ and $\cB_{\min}$ are indeed equivalent. We can then turn this
$1$-valued automaton into an unambiguous one (Figure~\ref{fig:ex2}-(b)),
using the construction of Proposition~\ref{pr-weber}.

This example is ``artificial''. For instance, we can 
get an equivalent two states unambiguous automaton from the max-plus one
only by deleting some transitions. This does not imply that there always
exists an unambiguous automaton that has a number of states less or equal to
the number of states of either the max-plus or the min-plus automaton.
We now give an example that enhances this point.

\medskip

Recall first that every max-plus or min-plus series over a one-letter alphabet is
unambiguous~\cite{BoKr,moll}.
We now make the following claim (the proof is not difficult): 
If $S$ is a max-plus rational series over the one-letter alphabet
$\{a\}$, and if the sequence $(\coef{S}{a^n})_{n\in \N}$ is periodic
of minimal period $p$, then the smallest 1-valued automaton
recognizing $S$ is of dimension $p$, and is deterministic. 

Let $p,q,r$, and $s$ be four distinct prime numbers. For $i\in \{p,q,r,s\}$,
define the series $S_i$ on $\{a\}^*$ by:
\[
\supp{S_i}=\{a^n\mid n=0\mod i\}, \quad \forall w \in \supp{S_i}, \ 
\coef{S_i}{w}=i\:.
\]
If $w$ is not in $\supp{S_i}$, set
$\coef{S_i}{w}=\mathbb{0}$ with the convention that $\mathbb{0}$
is neutral for both $\min$ and $\max$ and absorbing for $+$.
We then consider the series $T$ defined by:
\begin{eqnarray*}
\forall w\in a^*,&& \coef{T_1}{w}= \max(\coef{S_p}{w},\coef{S_q}{w}),
\quad \coef{T_2}{w} = \min(\coef{S_r}{w},\coef{S_s}{w})\\
&&  \coef{T}{w}=\coef{T_1}{w} + \coef{T_2}{w} \:.
\end{eqnarray*}
The series $T_1$ and $T_2$, and therefore $T$, are unambiguous, so
they belong to $\Zmax\Rat(a^*)\cap \Zmin\Rat(a^*)$. 
\begin{figure}[ht]
\[\epsfxsize=160pt \epsfbox{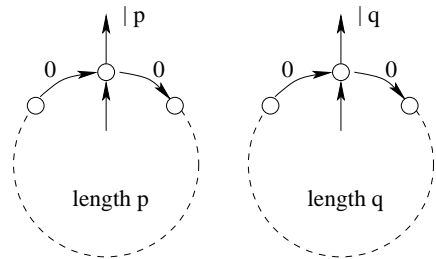} \]
\caption{A max-plus automaton recognizing $T_1$.}\label{fi-last}
\end{figure}
The series $T_1$ is recognized by the max-plus
automaton of dimension $(p+q)$ given in Figure \ref{fi-last}. 
A min-plus (and determinisic) automaton recognizing $T_1$ is the
following one (for $p<q$): 

States: $\{0,1,\dots, pq-1\}$; transitions: $i
\stackrel{a|0}{\longrightarrow} i+1 \mod pq$; initial state:
$\stackrel{|0}{\rightarrow} 0$; final states:
$ip\stackrel{|p}{\rightarrow}$ for $1 \leq i<q$, and
$jq\stackrel{|q}{\rightarrow}$ for $0\leq j<p$.

\medskip

And similarly for $T_2$, the small
automaton being the min-plus one. 
Therefore, the series $T$ is recognized by a max-plus automaton of
dimension $(p+q)rs$, and a min-plus one of dimension $pq(r+s)$. Now
observe that $(\coef{T}{a^n})_{n\in \N}$ is periodic of minimal period $pqrs$. 
Using the above claim, the smallest 1-valued (or unambiguous, or
deterministic) automaton recognizing $T$ is of dimension $pqrs$. 


\end{document}